%% file: conference_101719.tex
\documentclass[conference]{IEEEtran}
\IEEEoverridecommandlockouts
\input{Preliminaries.tex} 
\usepackage{filecontents}
\usepackage{nomencl}
\makenomenclature
\usepackage{etoolbox}
\renewcommand\nomgroup[1]{%
  \item[\bfseries
  \ifstrequal{#1}{A}{Sets and Indexes}{%
  \ifstrequal{#1}{B}{Parameters}{%
  \ifstrequal{#1}{C}{Variables}{}}}%
]}
\usepackage{graphicx}

\usepackage{cite}
\usepackage{amsmath,amssymb,amsfonts}
\usepackage{algorithmic}
\usepackage{multirow}
\usepackage[numbers]{natbib}
\usepackage[normalem]{ulem}
\usepackage{mathtools}
\begin{document}

\title{Liberalized market designs for district heating networks under the EMB3Rs platform\\
}

\author{\IEEEauthorblockN{1\textsuperscript{st} António Faria}
\IEEEauthorblockA{\textit{INESC TEC} \\
Porto, Portugal \\
antonio.s.faria@inesctec.pt}
\and
\IEEEauthorblockN{2\textsuperscript{nd} Tiago Soares}
\IEEEauthorblockA{\textit{INESC TEC} \\
Porto, Portugal \\
tiago.a.soares@inesctec.pt}
\and
\IEEEauthorblockN{3\textsuperscript{rd} José Maria Cunha}
\IEEEauthorblockA{\textit{INEGI} \\
Porto, Portugal \\
jmcunha@inegi.up.pt}
\and
\IEEEauthorblockN{4\textsuperscript{th} Zenaida Mourão}
\IEEEauthorblockA{\textit{INEGI} \\
Porto, Portugal \\
zmourao@inegi.up.pt}

}

\maketitle

\begin{abstract}
Current developments in heat pumps, supported by innovative business models, are driving several industry sectors to take a proactive role in future district heating and cooling networks in cities. For instance, supermarkets and data centers have been assessing the reuse of waste heat as an extra source for the district heating network, which would offset the additional investment in heat pumps. This innovative business model requires complete deregulation of the district heating market to allow industrial heat producers to provide waste heat as an additional source in the district heating network.

This work proposes the application of innovative market designs for district heating networks, inspired by new practices seen in the electricity sector. More precisely, pool and \gls{P2P} market designs are addressed, comparing centralized and decentralized market proposals. An illustrative case of a Nordic district heating network is used to assess the performance of each market design, as well as the potential revenue that different heat producers can obtain by participating in the market. An important conclusion of this work is that the proposed market designs are in line with the new trends, encouraging the inclusion of new excess heat recovery players in district heating networks.
\end{abstract}


\nomenclature[A]{$\Omega_n$}{Set of agents \textit{n}}
\nomenclature[A]{$\Omega_m$}{Set of agents \textit{m}}
\nomenclature[A]{$\Omega_c$}{Set of consumers \textit{c}}
\nomenclature[A]{$\Omega_p$}{Set of producers \textit{p}}
\nomenclature[A]{$\Omega_{k}$}{Set of communities \textit{k}}
\nomenclature[A]{$\Omega_{I_{ n,m}}$}{Set of pipes in the path between agents $n$,$m$}
\nomenclature[A]{$n$}{Agents index}
\nomenclature[A]{$m$}{Agents index}
\nomenclature[A]{$k$}{Community index}
\nomenclature[A]{$i$}{Pipe index}
\nomenclature[A]{$t$}{Time period index}

\nomenclature[B]{$C_n$}{Price offer by agent $n$}
\nomenclature[B]{$\underline{P}_n , \overline{P}_n$}{Lower and upper  bounds of agent \textit{n}}
\nomenclature[B]{$C_{n,m}$}{Product differentiation cost  applied to the trade between agents \textit{n} and \textit{m}}
\nomenclature[B]{$c_{n,m}$}{Initial penalty between agents $n$ and $m$}
\nomenclature[B]{$d_{i,n,m}$}{Distance of pipeline $i$ in the path between agents $n$ and $m$}
\nomenclature[B]{$l_{n,m}$}{Thermal losses of pipeline $i$ in the path between agents $n$ and $m$}
\nomenclature[B]{$D_{i,n,m}$}{Influence of agents $n$ and $m$ in the total heat flow in pipe $i$ (\%)}
\nomenclature[B]{$C_{n,k}$}{Price offer by agent $n$ from community $k$}
\nomenclature[B]{$c_{exp,k}$}{Cost of heat export}
\nomenclature[B]{$c_{imp,k}$}{Cost of heat import}
\nomenclature[B]{$E_{n}$}{$CO_2$ Signals Coefficient by agent $n$}
\nomenclature[B]{$Tot Dist$}{Total pipeline distance in the network}
\nomenclature[B]{$Tot Loss$}{Total pipeline losses in the network}

\nomenclature[C]{$P_{n}$}{Total heat production/consumption of agent \textit{n}}
\nomenclature[C]{$P_{n,m}$}{Bilateral trade between agents \textit{n} and \textit{m}}
\nomenclature[C]{$P_{n,k}$}{Total heat production/consumption of agent \textit{n} from community $k$}
\nomenclature[C]{$q_{exp,k}$}{Heat exported by a community $k$}
\nomenclature[C]{$q_{imp,k}$}{Heat imported by a community $k$}
\nomenclature[C]{$P_{k,k'}$}{Bilateral trade between communities \textit{k} and \textit{k'}}
\nomenclature[C]{$q_{n,k'}$}{Internal trade by agent $n$ at the community $k$}
\nomenclature[C]{$\alpha_{n,k'}$}{Imported heat by agent $n$ at the community $k$}
\nomenclature[C]{$\beta_{n,k'}$}{Exported heat by agent $n$ at the community $k$}
\nomenclature[C]{$Participation_{p,t}$}{Binary variable indicating market participation}

\printnomenclature

\section{Introduction}
Over the years, \gls{DHC} systems have been proliferating in many countries \cite{Mathiesen2019}. In Denmark, according to EUROHEAT \& POWER, 65\% of citizens were served by \glspl{DHN} in 2017, accounting for more than 30 000 km of pipelines in \glspl{DHN}. 
Most European \gls{DHC} systems follow a monopolistic approach due to heat demand sparsity, the market power of a single generating unit that often owns the \gls{DHN}, the lack of \gls{DHN} linking all possible customers, and long-term return on investment. These reasons pull back new investors and market liberalization, which could foster the reuse of waste heat as an extra source in \glspl{DHN} \cite{Persson2011,Buffa2019}.
In fact, \gls{DHN} is a natural monopoly due to the large infrastructure and operation costs, concerning the production and distribution of heating and cooling. Therefore, the heat production plants and the network are commonly owned, operated and managed by the same company, which is the main obstacle to the complete liberalization of the system \cite{Magnusson2016}. Overall, \gls{DHC} systems are heavily regulated and price competitiveness for consumers is disregarded.

Nevertheless, governments (through energy regulators and policymakers) are enforcing the liberalization of heat markets (similar to what happened in the power system), as it becomes easier to monitor the whole process of energy systems, aiming to drag the prices down through competition, once the energy providers are competing with each other, leading to economic benefits for consumers \cite{Westin2002,Soderholm2011,Liu2019,Zhang2013, Wissner2014}. Therefore, \gls{DHC} market liberalization is gaining momentum in some European countries, aiming to replicate and adapt the good experience with electricity markets, bringing their capacity to improve system efficiency \cite{Gebremedhin2004,Liu2019,Gulzar2015}. This disruptive paradigm shift will increase competitiveness through the inclusion of new players in the system. That is, several agents from different industry sectors can play an active role in the \gls{DHC} market by buying and selling energy from different sources, increasing competitiveness and bringing financial benefits to everyone involved \cite{Soderholm2011,Burger2019}. The authors in \cite{Gebremedhin2004, Marinova2008} present case studies suggesting that a large amount of heat demand can be supplied by industries, e.g., by supplying waste/excess heat of industry processes to neighbouring consumers. Similarly, the authors in \cite{Syri2015, Brand2014} also demonstrate the benefits that external producers (taking advantage of heat pumps, waste heat and renewable heat technology) bring to the \gls{DHC} system if they supply their excess heat to the \gls{DHN}. The results would be advantageous for all parties, bringing economic and environmental gains. On the other hand, the works in \cite{Liu2016, Huang2017, Lu2018} show the benefits of the synergies between the power and \gls{DHC} systems, modeling centralized dispatches to improve the efficiency of the entire energy system. In addition, consumers can also play an active role in the \gls{DHC} system, providing demand flexibility in response to dynamic tariffs, thereby improving market competition \cite{Dominkovic2018,Li2019,Djorup2020,Bhattacharya2016}.

\gls{DHC} markets inspired by the electricity sector, applying conventional market designs and approaches, are growing \cite{Pazeraite2013,Gulzar2015}. An example of a running \gls{DHC} market is the Open District Heating project \cite{opendistrict}, operating at Stockholm's \gls{DHN}, which encourages industrial businesses to sell their excess heat to the \gls{DHN} at a uniform price cleared in the proposed day-ahead heating market.

In addition, innovative market ideas to increase competitiveness in the \gls{DHN} are emerging in the literature \cite{Li2015,Moshkin2016, Valeriy2019}. One of them is the adaption of the sharing economy principle to industries and small-scale production units to supply surplus heat to the \gls{DHN} \cite{Marinova2008,Karlsson2009}. In this regard, different consumer-centric market designs, adapted from the power system, are expected to be replicated to the \gls{DHC} system, allowing these new market participants to inject heat in the \gls{DHN} and get extra revenue.
In order to assess several options and assumptions for the best market design to apply in existing and new \glspl{DHN}, a brand new platform (EMB3Rs) is being developed \cite{embers}. This platform will empower different stakeholders (e.g., utility companies, municipalities, \gls{DHN} operators, excess heating producers, among other entities) to simulate distinct market designs that can be applied to current and future \glspl{DHN}.

In this context, this work contributes to the literature and to the EMB3Rs platform, modelling distinct market models for the negotiation of heat in \glspl{DHN} considering a competitive environment. More precisely, three distinct market designs are modelled and compared, namely, the pool-based, the peer-to-peer (P2P), and the community-based market designs. The markets are adapted from the current and future trends in electricity markets. Additionally, consumers preferences (e.g., distance, losses and $CO_2$) through product differentiation are applied to the P2P market design, enabling consumers to choose sources they prefer to be provided from. An illustrative \gls{DHN} based on Nordic countries is used to test the applicability of the proposed solution. The main contributions of the present work are fourfold:

\begin{itemize}
    \item To implement, analyze and compare, different market models in the EMB3Rs platform;
    \item To model new market designs for heat exchange in the \gls{DHN}, namely, the pool-based, P2P, and community-based market designs;
    \item To explore competitiveness in \gls{DHC} markets, enabling industrial businesses with excess heat recovery systems to inject excess heat in the DHN;
    \item To improve market options for consumers by introducing product differentiation in the P2P market design.
\end{itemize}

In addition to this introductory section, this paper is organized as follows. Section two describes the EMB3Rs platform for the simulation of different \gls{DHC} market designs. Section three presents the detailed mathematical models of the proposed market designs. Section four assesses the proposed market models considering an illustrative case of Nordic \glspl{DHN}, while section five gathers the conclusions of the study.

\section{EMB3Rs Platform for \gls{DHC} Market Simulation}
This section provides an overview of the EMB3Rs platform that will incorporate current and new market designs, adapted to the context of \gls{DHC} systems. In addition, it provides a brief review of the actual situation of the \gls{DHC} markets in the Nordic countries. 

\subsection{Current \gls{DHC} Market Situation in Nordic Countries}

The current situation of \gls{DHC} markets varies on a country basis, as the deregulation of \gls{DHC} systems has been carried out in different ways \cite{climatex}. In Denmark, the \gls{DHN} is still a natural monopoly, as the network and heating plants are mostly owned by energy companies, municipalities or consumer cooperatives. The regulation dictates that the heat supply works under non-profit rules, which means that the supplier must provide heat to consumers at marginal cost. This non-profit rule benefit everyone, as any profits are distributed to consumers to reduce costs \cite{DanishEnergyAgency2016}. In this case, industries with excess heat are encouraged to self-consume and only then to sell excess heat to the market, since the sale of excess heat comes with a tax to prioritize energy efficiency \cite{climatex}.

Similarly to Denmark, \glspl{DHN} are also heavily regulated in Norway. \glspl{DHN} are mostly private and municipal owned, with mandatory connections to consumers decided by the municipalities, while the operator is forced to expand the network \cite{climatex}. The energy price from different producers are set on a competitive market, but prices for consumers with mandatory connections are regulated and cannot exceed the price of electric heating within the supply area \cite{Aanensen2014}. Alternatively, consumers without mandatory connections are free to choose their heating source (e.g., electric heating or heat pump), so the supply price will follow the electricity price \cite{Aanensen2014, Hawkey2014}.

In contrast, Sweden was one of the first European countries to deregulate the heating market, however, that deregulation was not as robust as expected. According to \cite{ Aberg2016}, the prices of the different Swedish utility companies are not similar, meaning that these companies behave as price-makers. The costs are related to heating production and \gls{DHN} operation, while what was expected was marginal-based pricing. On the other hand, Finish utility companies have a monopoly on certain \gls{DHN}s. Costumers have no open market to select their \gls{DHC} utility \cite{Paiho2016}. Some Finish companies have been trying to change this paradigm, i.e., offering seasonal tariffs, but these measures also do not shape the fair price for customers \cite{Dominkovic2018}. For further details on the situation of \glspl{DHC} systems in European countries, interested readers are referred to \cite{climatex, Werner2017}. 

Nonetheless, the transition to sustainable, efficient and competitive markets is unavoidable and future \gls{DHC} markets will require new market approaches suitable to the integration of renewable energy sources in \glspl{DHN} \cite{Sorknaes2020}.

\subsection{EMB3Rs Platform Overview}
The EMB3Rs platform has been designed to assess the reuse and trade of excess thermal energy in a holistic perspective within an industrial process, energy system environment, or in an \gls{DHN} under regulated and liberalized market environment \cite{emb3rs_2}. The platform empowers industrial users and stakeholders to investigate the revenue potential of using industrial excess heat and cold as an energy resource, based on the simulation of supply-demand scenarios. Therefore, the platform simulates multiple business and market models, proposing innovative solutions in the sector. 

From the large variety of options, users can: (\textit{i}) map new and existing supply and demand users with geographic relevancy and enable their interlink;
(\textit{ii}) assess costs and benefits related to the excess heat and cold utilization routes, considering existing and new network infrastructure (e.g., \gls{DHN}); (\textit{iii}) explore and assess the feasibility of new technology and business scenarios; and (\textit{iv}) compare and analyze distinct market models applied to the \gls{DHN} to dynamically create new business models and identify potential benefits and barriers under specific regulatory framework conditions. 

The integration of a dedicated market module in EMB3Rs platform allows users to perform market analysis considering multiple existing market designs. Therefore, users can create, test and validate different market structures for selling and buying energy in the \gls{DHN}, identifying barriers and risks, as well as regulatory framework conditions required to ensure that the implementation of such market solutions are economically feasible. That is, the market analysis enable users (e.g. industries, supermarkets and data centers) to estimate potential revenues from selling excess heat and cold. This is especially important for users who have invested (or are considering investing) in waste heat recovery technology to assess the potential economic and environmental savings of their investment. 



\subsection{Market Approach for Heat Exchange}
On the EMB3Rs platform, users must be able to explore different market designs, from centralized to the decentralized designs, allowing them to analyze the best market framework for their interests, which can be economic, environmental or social.

In this regard, three distinct market designs are adapted in the present work to be included in the EMB3Rs platform. The conventional pool market, the innovative \gls{P2P} and community-based market designs are addressed to ensure that the platform's users (e.g., industries, supermarkets and data centers) can assess their business models under different levels of market decentralization for the exchange of thermal energy in \gls{DHN}s. All the three market designs are inspired in the electricity sector, and therefore, need to be adapted to the underlying characteristics of \gls{DHC} systems. 

The pool market follows a systemic perspective of the whole market by applying the merit order mechanism and performing the intersection of production and demand curves. This mechanism, known as uniform price, results in a market clearing price that is used for the settlement of producers and consumers. That is, each producer and consumer scheduled in the market will receive and pay for the energy at the market clearing price, respectively.

In contrast, consumer-centric market designs (such as \gls{P2P} and community-based market designs) follow a more decentralized and consumer-focused perspective. The P2P market enables producers and consumers to exchange energy directly with each other, subject to certain specific conditions defined by consumers. In this market design, no central facilitator is needed to verify energy exchanges. On the other hand, the community-based market requires the use of a central entity that coordinates energy exchanges within the energy community, well as the imports and exports to other energy communities and \gls{DHN} players. It worth mention that these kind of decentralized markets can empower consumers and prosumers to play a more active role in the \gls{DHN}. For instance, local supermarkets are emerging thermal prosumers that can provide and consume heat in different hours, making them a flexible player to reuse excess heat and even selling surplus heat to other consumers in the \gls{DHN}.     

\input{Heat_P2P_markets}

\section{Case study}

In this section, a case study is presented considering an illustrative Nordic \gls{DHN} with several producers and consumers. This illustrative example has been developed to assess different market designs on the EMB3Rs platform.
All the input data and results of this study, including demand and supplier offers for an entire year (from April 2018 to March 2019) are available at Mendeley Data \cite{dataset_men}.

\subsection{Case Characterization}
A \gls{DHN} has been built considering several producers and consumers with different characteristics and patterns. 
\begin{figure}[!h]
        \begin{center}
        \includegraphics[width=0.5\textwidth]{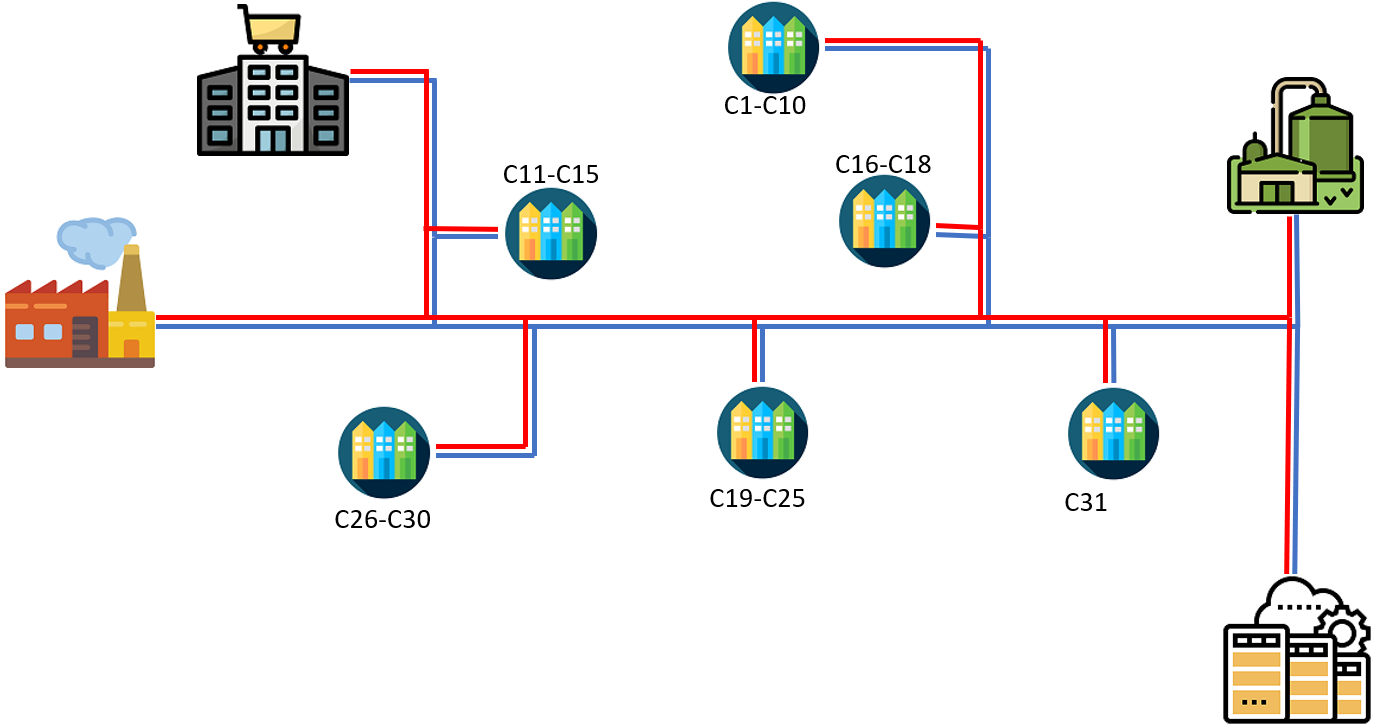}
        \caption{Illustrative district heating network.}
        \label{fig:dhn}
    \end{center}
\end{figure}

Note that the \gls{DHN} must ensure that the temperature is within the levels required by the heating demand, and that the flow rates in the \gls{DHN} must be kept at a reasonable low level in order to avoid water velocities. To this end, it is assumed that this \gls{DHN} operates similarly to most Danish \glspl{DHN}, which work within annual averages temperatures of 77.6$^\circ$C supply and 43.1$^\circ$C return \cite{Gadd2014}.

Figure \ref{fig:dhn} shows the schematic diagram of the \gls{DHN}, where 31 row houses and 4 potential producers are considered. The consumption of 31 row houses for a entire year (from April 2018 to March 2019) has been generated considering a typical demand pattern taken from \cite{BrandThesis2014}. The price that the row houses are willing to pay for the demand in the market follows a normal distribution, in which the base price is the heat tariff in Copenhagen, Denmark \cite{hofor}. In order to suppress basic consumption needs, at least 70\% of the heat demand of each house must be supplied at all periods.

A 15 kW industrial ammonia heat pump is located in the \gls{DHN} and can provide heat at some time of the day at a certain cost. The heat pump generation profile considers a constant Coefficient of Performance (COP) of 4.8, providing hot water via a heat exchanger at 80$^\circ$C, based on \cite{Ommen2013,Fukano2011}. The cost curve of the heat pump is based on the electricity spot price in 2018 and 2019 in DK2 area in Denmark, taken from \cite{energi}. 

In addition, a 0.4 MW data center is included in the \gls{DHN}. Commonly, data centers follow a relatively constant pattern of excess heat recovery to inject into the \gls{DHN}, although the temperature of their excess heat from the condenser cooling towers is usually between 35$^\circ$C and 45$^\circ$C. Thus, an industrial ammonia heat pump, similar to the one referred above, would be required to upgrade its heat to inject into the \gls{DHN}. This data center has been modeled producing 71,6 kWh on average, in which the calculus for the heat recovery profile is based on \cite{datacenter,WAHLROOS20171228}. To this value, it would be added the energy used in the ammonia compressor. The cost curve for the data center sell recovered heat energy in the \gls{DHN} has been modeled following a normal distribution and the monthly excess heat procurement costs presented in \cite{WAHLROOS20171228}. 
A \gls{CHP} unit is included in the \gls{DHN} being the main producer in the system. This CHP is designed to provide the entire consumption of the system, being therefore the most expensive generation resource. The cost curve for a entire year follows the behavior of the natural gas spot price for years 2018 and 2019, available in \cite{henry_hub}. Note that the prices were normalized for the Nordic context. 

Besides this, a supermarket with heat pump technology is included in the system behaving like a prosumer. That is, the supermarket may consume heat from the \gls{DHN} or inject recovered heat into the \gls{DHN}, taking into account the hour of the day and the outdoor temperature. The generation and consumption profile depends on the outdoor temperature. It has been considered the outdoor temperature in Copenhagen for the entire year (April 2018 to March 2019), available at \cite{weather}. Then, the prosumer profile of the supermarket is determined following a typical COP (around 3.0) for heat recovery in supermarkets, and a typical supermarket consumption pattern, detailed in \cite{Karampour2014}. The cost curve for the supermarket to inject recovered heat in the \gls{DHN} depends on the outdoor temperature and is based on \cite{Id2018}. 

It is noteworthy that different market designs may require the use of different data or configurations. For example, the community-based market design requires the configuration of the energy community, that is, who are the community members.
For the community-based market, two communities were created, based on the aforementioned energy resources, namely: 
\begin{itemize}
    \item Community 1: Data Center and all consumers from 19 to 31;
    \item Community 2: Supermarket, Heat Hump, and consumers from 1 to 18.
\end{itemize}

Regarding the P2P market model via product differentiation, the required data were retrieved based on the THERMOS project tool \cite{thermos}. This tool is able to provide the distance (Table \ref{tab:distance}) and nominal losses (Table \ref{tab:losses}) between agents, based on the supply and return temperatures, and on the maximum heat flow in the pipelines.

 The $CO_2$ signals for the \gls{CHP} were obtained from \cite{EIB2018}, while for technologies that rely on the electricity mix were retrieved from \cite{co2_intensity} considering the Nordic zone. Table \ref{tab:co2_signals} presents the $CO_2$ signals for all heat producers.

\begin{table}[]
\caption{DHN distance between agents.}\label{tab:distance}
\resizebox{\columnwidth}{!}{%
\begin{tabular}{@{}ccccc@{}}

        & \multicolumn{4}{c}{Distance (m)} \\ 
Agent   & CHP    & Supermarket     & Data Center     & Heat Pump     \\
C1-C10  & 266,24  & 181,25 & 206,15 & 174,96 \\
C11-C15 & 190,76  & 20,47  & 168,58 & 199,06 \\
C16-C18 & 228,66  & 143,67 & 230,27 & 137,38 \\
C19-C25 & 175,25  & 90,26  & 158,21 & 127,01 \\
C26-C30 & 196,37  & 111,38 & 224,52 & 193,31 \\
C31     & 259,32  & 174,33 & 122,23 & 94,28  \\
SM      & 201,37  & -      & 240,87 & 209,67 \\ 
\end{tabular}
}
\end{table}

\begin{table}[]
\caption{DHN nominal losses between agents.}\label{tab:losses}
\resizebox{\columnwidth}{!}{%
\begin{tabular}{@{}ccccc@{}}

        & \multicolumn{4}{c}{Losses  (W/m)} \\ 
Agent   & CHP   & Supermarket     & Data Center     & Heat Pump     \\
C1-C10  & 17,31  & 16,40  & 17,31  & 14,02  \\
C11-C15 & 18,35  & 17,23  & 16,83  & 17,43  \\
C16-C18 & 17,90  & 17,12  & 17,73  & 17,58  \\
C19-C25 & 18,10  & 18,01  & 17,78  & 17,49  \\
C26-C30 & 17,24  & 16,51  & 17,41  & 16,43  \\
C31     & 17,39  & 16,99  & 17,05  & 16,66  \\
SM      & 18,64  & -      & 16,87  & 17,86  \\ 
\end{tabular}
}
\end{table}

\begin{table}[]
\caption{$CO_2$ emissions by heat producer.}\label{tab:co2_signals}
\begin{tabular}{@{}cccc@{}}

\multicolumn{4}{c}{$CO_2$ Signals (g/kW)}       \\ 
CHP & Supermarket & Data Center & Heat Pump \\
225  & 225         & 166.1         & 34.6     \\ 
\end{tabular}
\end{table}

\subsection{Results}

This section presents the main results and indicators for comparing the different market designs. All simulations were performed for an entire year of market operation.

\subsubsection{General Results}

Table \ref{tab:agents_revenue} presents the social welfare and the revenue achieved by each agent over the simulated year. For the pool market, the achieved results are the same as the Full P2P, so these are not discussed in detail. As expected, the Full P2P market design is the one presenting the best solution, since there are no limitations on heat exchanges between agents, opposite to what happens in P2P with product differentiation where penalties (consumer preferences) are considered. Note that social welfare represents the objective function without penalties, i.e., once the objective is defined, the penalties are removed and all heat transactions are kept. Within the P2P markets, the P2P with distance as product differentiation (P2P Distance) is the one achieving the lowest social welfare (65,9\% compared to Full P2P), since it is the one that most penalizes the transactions between agents. P2P CO2 is the one reaching the social welfare closest to the Full P2P (more than 99.8\%). The Full P2P and the community-based are the market models supplying more load, reaching 90\% of the total load demand. Other models have a smaller delivery capacity and the minimum is reached for the P2P Distance where only 70\% of the entire load demand is met. Although the community performs the poorest social welfare (63\% compared to the Full P2P), it is worth stressing that it is the market that allocates the most load. In terms of heat production, the \gls{CHP} and the data center are the ones producing the most heat throughout the year. The \gls{CHP} has the largest thermal energy producing capacity and is the most expensive resource. Thus, it is often used to cover the remaining energy demand, which other producers cannot cover. On the other hand, the high dispatch of the data center is related to its high nominal capacity and low bid price offered in the market. The \gls{CHP} shows a drop of about 45 \% in production at P2P Distance when compared to the Full P2P, which is linked to the fact that it is the producer that is more distant from the consumers.

It is worth mentioning that the heat pump reaches high dispatched heat levels and, consequently, high revenue in the P2P Distance and Community-based markets. The heat pump is located very close to the consumption points, which helps to explain the heat pump performance in the market design that considers the distance between agents. With respect to the community-based, the heat pump results are related to the community structure. The heat pump is a member of Community 2, where only the supermarket compete to meet the demand. As the supermarket behaves as a prosumer, the heat pump or imported heat are often the only available heat sources for that community, leading to a higher market share for the heat pump. As the heat pump and the data center are the two sources with the lowest $CO_2$ emissions, these are also the only agents presenting an increase in the heat supplied (1.8\% and 6.7\%, respectively), when comparing the P2P $CO_2$ with the Full P2P.

\begin{table}[]
\caption{Agents' revenue by market design}\label{tab:agents_revenue}
\resizebox{\columnwidth}{!}{%
\begin{tabular}{@{}cccccc@{}}

\multicolumn{6}{c}{Revenue (€)}                                             \\ 
            & Full P2P     & P2P Distance & P2P Losses & P2P $CO_2$ & Community \\ 
Social Welfare  & 175250 & 115560      & 166422    & 175040 & 110407       \\
CHP        & 89328   & 69179        & 78094      & 85115   & 185057        \\
Supermarket & 5615    & 6162         & 5813       & 5352    & 6093          \\
Data Center & 85090   & 77614        & 84670      & 86931   & 77452         \\
Heat Pump   & 6610    & 13413        & 5338       & 7007    & 14113         \\
Load        & 361893  & 281928       & 340338    & 359446  & 366479        \\ 
\end{tabular}
}
\end{table}

\begin{table}[]
\caption{Agents' dispatched heat by market design}\label{tab:dispatched_heat}
\resizebox{\columnwidth}{!}{%
\begin{tabular}{@{}cccccc@{}}

\multicolumn{6}{c}{Dispatched Heat (kW)}                                   \\ 
            & Full P2P    & P2P Distance & P2P Losses & P2P $CO_2$ & Community \\
Load  & 682941 & 532850       & 642078     & 678188  & 687215       \\
CHP        & 217191 & 120623       & 180546     & 205486  & 275674       \\
Supermarket & 39937  & 43255        & 43255      & 38173   & 42758         \\
Data Center & 411472 & 338954       & 408897     & 419155  & 336219       \\
Heat Pump   & 14341  & 30018        & 11522      & 15372   & 32564         \\ 
\end{tabular}
}
\end{table}

\subsubsection{Average Dispatched Heat and Successful Participation in the Market}

In addition to the general results, two key performance indicators (namely, the \gls{ADH} and the \gls{SPM}), were introduced. \gls{ADH} represents the amount of heat that is dispatched from a source on average, i.e., the mean percentage of dispatched heat from the total capacity of the source. The values are presented in percentage (\%) and determined through: 

\begin{equation}\label{eq:avg_disp}
    ADH(n{)}=\frac{\sum_{t=1}^{T}\frac{P_{n,t}^{}}{\overline{P}_{n,t}}}{T},\forall n\in\left\{\Omega_p\right\}
\end{equation}
where $P_{n,t}$ represents the heat dispatched by source $n$ in time period $t$ and $\overline{P}_{n,t}$ represents the maximum capacity of source $n$ in time period $t$.

Regarding the \gls{SPM}, it indicates the level of participation by an agent $n$ in the market, which is given by:

\begin{equation}
    SPM(n{)}=\frac{\sum_{t=1}^{T}{Participation_{(n,t{)}}}}{T}\times100,\forall n\in\left\{\Omega_p\right\}
\end{equation}
where $Participation_{n,t}$ is a binary variable indicating whether a source $n$ is or not dispatched in the market, in time frame $t$.

In addition to the annual results, seasonal results are also presented, once the sources and loads have seasonal behaviors. As one can see in Table \ref{tab:adg}, the heat dispatched is generally higher in the winter, which is linked to lower external temperatures, hence larger levels of heat demand are required. However, the \gls{CHP} presents lower \gls{ADH} in the winter when compared to the summer period. This is connected to the higher bidding prices offered by this resource in that period of the year, which enhances other resources participation in the market. Also note that the supermarket is the resource with the highest \gls{ADH}, being fully dispatched most of the time. It is also noteworthy that the heat pump is less dispatched in the summer than in the winter, not only due to the increase of the bid offer, but also due to the lower production capacity during this season.

\begin{table}[]
\caption{Annual and seasonal index of average dispatched heat for each heat producer and market design.}\label{tab:adg}
\resizebox{\columnwidth}{!}{%
\begin{tabular}{@{}ccccc@{}}

              & \multicolumn{4}{c}{Year}                            \\ 
              & CHP   & Supermarket   & Data Center   & Heat Pump  \\
Full P2P           & 72\%   & 97\%          & 62\%          & 25\%       \\
P2P Distance  & 71\%   & 100\%         & 51\%          & 64\%       \\
P2P Losses    & 71\%   & 98\%          & 61\%          & 14\%       \\
P2P $CO_2$       & 72\%   & 96\%          & 63\%          & 28\%       \\
Community & 30\%   & 100\%         & 51\%          & 91\%       \\ 
              & \multicolumn{4}{c}{Summer}                          \\ 
              & CHP   & Supermarket   & Data Center   & Heat Pump  \\
Full P2P           & 84\%   & 97\%          & 48\%          & 1\%        \\
P2P Distance  & 83\%   & 100\%         & 29\%          & 36\%       \\
P2P Losses    & 87\%   & 98\%          & 48\%          & 1\%        \\
P2P $CO_2$       & 84\%   & 96\%          & 49\%          & 4\%        \\
P2P Community & 34\%   & 100\%         & 31\%          & 92\%       \\ 
              & \multicolumn{4}{c}{Winter}                          \\ 
              & CHP   & Supermarket   & Data Center   & Heat Pump  \\
Full P2P           & 60\%   & 98\%          & 76\%          & 50\%       \\
P2P Distance  & 58\%   & 100\%         & 73\%          & 92\%       \\
P2P Losses    & 54\%   & 98\%          & 76\%          & 28\%       \\
P2P $CO_2$       & 60\%   & 97\%          & 77\%          & 53\%       \\
Community & 26\%   & 100\%         & 71\%          & 90\%       \\ 
\end{tabular}
}
\end{table}

Regarding the \gls{SPM} indicator, the results clearly point to a high successful participation of the supermarket and data center in all market designs. When it comes to the data center, these results are justified by its steady heat production and low offer price, being one of the first sources that all consumers want to exchange with. It is important to highlight the contrast exhibited between \gls{SPM} and \gls{ADH} in relation to the \gls{CHP}, since in the summer there is less heat demand that can be met by other agents with better offers, thus reducing this agent overall participation.

\begin{table}[]
\caption{Annual and seasonal index of successful participation in the market for each heat producer and market design.}\label{tab:spm}
\resizebox{\columnwidth}{!}{
\begin{tabular}{@{}ccccc@{}}

              & \multicolumn{4}{c}{Year}                                     \\ 
              & CHP     & Supermarket     & Data Center     & Heat Pump     \\
Full P2P           & 36\%     & 91\%            & 89\%            & 26\%          \\
P2P Distance  & 61\%     & 100\%           & 100\%           & 64\%          \\
P2P Losses    & 37\%     & 99\%            & 100\%           & 16\%          \\
P2P $CO_2$       & 35\%     & 88\%            & 90\%            & 28\%          \\
Community & 81\%     & 100\%           & 100\%           & 92\%          \\ 
              & \multicolumn{4}{c}{Summer}                                   \\ 
              & CHP     & Supermarket     & Data Center     & Heat Pump     \\
Full P2P           & 13\%     & 83\%            & 93\%            & 1\%           \\
P2P Distance  & 56\%     & 100\%           & 99\%            & 37\%          \\
P2P Losses    & 15\%     & 99\%            & 100\%           & 1\%           \\
P2P $CO_2$       & 12\%     & 75\%            & 95\%            & 4\%           \\
Community & 71\%     & 100\%           & 100\%           & 93\%          \\ 
              & \multicolumn{4}{c}{Winter}                                   \\ 
              & CHP     & Supermarket     & Data Center     & Heat Pump     \\
Full P2P           & 60\%     & 95\%            & 85\%            & 51\%          \\
P2P Distance  & 66\%     & 100\%           & 100\%           & 93\%          \\
P2P Losses    & 59\%     & 98\%            & 100\%           & 31\%          \\
P2P $CO_2$       & 59\%     & 93\%            & 86\%            & 54\%          \\
Community & 91\%     & 100\%           & 100\%           & 91\%          \\ 
\end{tabular}
}
\end{table}

\subsubsection{Fairness Indicators}

Fairness indexes are also assessed in this work. The methodology of \cite{Moret2019,Jain1998} was followed to evaluate the resource allocation in each market design. These indicators are not meant to measure quantities, but rather to assess the relationships between the different agents and the impact that each of them brings to the whole system. To do so, \gls{QoS}, \gls{QoE} and \gls{MiM} were determined. \gls{QoS} represents how all the agents impact the heat distribution in the system, i.e., if all involved agents trade the same amount of heat, then the \gls{QoS} would be equal to 100\%. This index assesses the equilibrium in the system. \gls{QoE} points out the consumer satisfaction related to the heating price when trading with other agents. The \gls{MiM} indicator stands for the fairness in the prosumers and consumers field, where the ratio between the minimum and maximum values for each time period is calculated. If all the consumers trade the same amount of heat, then this index equals 100\%. Table \ref{tab:fairness_indicators} gathers the fairness indicators results.

As one can see, in general, the market modules present a \gls{QoS} around 20\%, meaning that there are agents with larger capacities when compared to other. This discrepancy leads to lower levels of \gls{QoS}. When looking at community 2, this index is even lower which is related to the heat pump impact in this community. For most of the year, this player is in charge of supplying the whole community, creating a huge impact, attracting a large part of the exchange within the community. The \gls{QoE}, related to the user viewpoint, presents similar values for all P2P designs. When analysing the communities, these values are substantially lower, due to the fewer competitiveness existing in each community. Therefore, agents are compelled to exchange with players who do not offer prices as favorable as their competitors at certain times, as in the P2P market models. The low values presented by \gls{MiM} point to the significant difference between the heat values that are exchanged among the different agents.

\begin{table}[]
\caption{Fairness indicators for each market model}\label{tab:fairness_indicators}
\begin{tabular}{ccccccc}
                           & \multicolumn{2}{c}{QoS}  & \multicolumn{2}{c}{QoE}  & \multicolumn{2}{c}{MiM} \\
Full P2P                   & \multicolumn{2}{c}{21\%} & \multicolumn{2}{c}{78\%} & \multicolumn{2}{c}{4\%} \\
P2P Distance               & \multicolumn{2}{c}{17\%} & \multicolumn{2}{c}{83\%} & \multicolumn{2}{c}{4\%} \\
P2P Losses                 & \multicolumn{2}{c}{21\%} & \multicolumn{2}{c}{79\%} & \multicolumn{2}{c}{5\%} \\
P2P CO2                    & \multicolumn{2}{c}{20\%} & \multicolumn{2}{c}{79\%} & \multicolumn{2}{c}{4\%} \\
\multirow{2}{*}{Community} & Com 1       & Com 2      & Com 1       & Com 2      & Com 1      & Com 2      \\
                           & 26\%        & 14\%       & 48\%        & 23\%       & 2\%        & 16\%      
\end{tabular}
\end{table}



\subsubsection{Supermarket Individual Analysis}

\begin{figure}[!h]
        \begin{center}
        \includegraphics[width=0.50\textwidth]{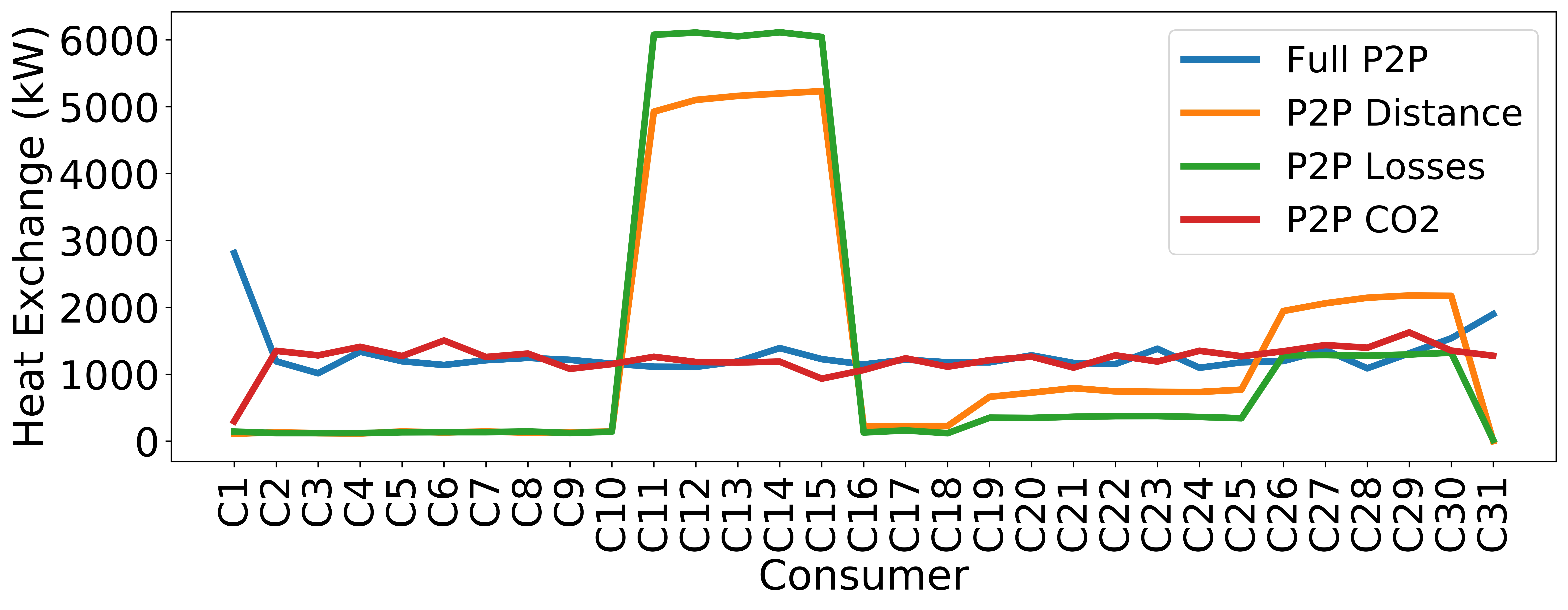}
        \caption{Cumulative annual heat exchange of the supermarket as a heat producer in the P2P designs.}
        \label{fig:supermarket_gen}
    \end{center}
\end{figure}

The supermarket is the only prosumer in the system, which means that it is the only player capable of behaving as a producer or consumer in different periods of time, being important to analyze its individual trades with other peers. When the supermarket is behaving as a producer, it is able to sell heat to the loads. Figure \ref{fig:supermarket_gen} depicts the cumulative heat trade over the year between the supermarket and the loads for each of the considered \gls{P2P} market designs. More precisely, figure \ref{fig:supermarket_gen} points to a steady supply to all consumers by the supermarket in the Full \gls{P2P} design, which was expected, since there are no preference constraints for any heat consumer. On the other hand, the product differentiation effect is clear in the \gls{P2P} Distance and P2P Losses, since consumer preferences (namely, distance and losses) encourage trading with closest peers. Thus, the consumers (C11-C15) are strongly encouraged to trade with the supermarket, as it is one of the closest producers. In fact, most of the supermarket heat production goes directly to these consumers (about 59.2\% and 73.9\% for \gls{P2P} Distance and P2P Losses, respectively), supplying other consumers with residual heat, or not at all.  In the \gls{P2P} considering the $CO_2$ signals, there are no major fluctuations once the $CO_2$ emissions value of the supermarket (225 g/kW) is similar to that of the \gls{CHP} and Data Center, and much higher than that of the Heat Pump. In this way, the differentiation criterion is minimal relative to the \gls{CHP} and Data Center with consumers giving priority to trade with the Heat Pump. More precisely, as both the supermarket and the Heat Pump have a low capacity to influence the system, the changes in the exchanges between the supermarket and the consumers are relatively small compared to the Full \gls{P2P} market design.

\begin{figure}[!h]
        \begin{center}
        \includegraphics[width=0.50\textwidth]{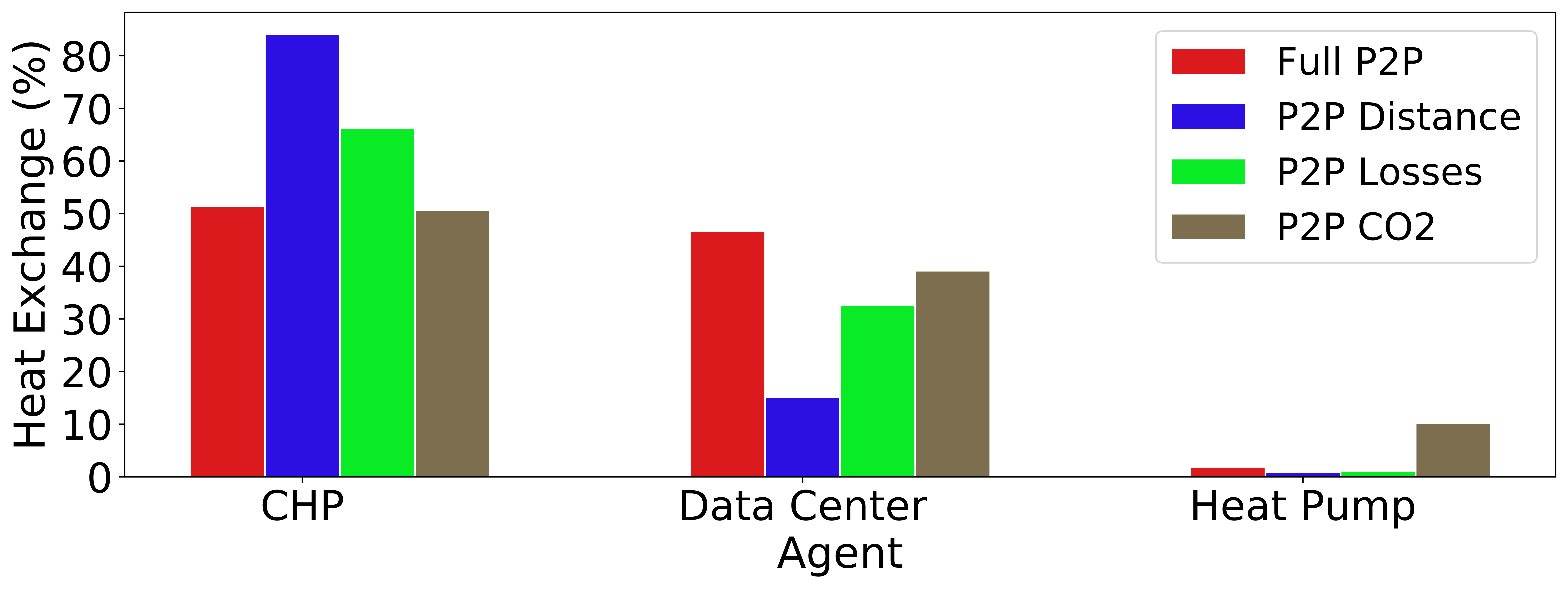}
        \caption{Cumulative annual heat exchange of the supermarket as a heat consumer in the P2P designs.}
        \label{fig:supermarket_dem}
    \end{center}
\end{figure}

Notwithstanding, there are periods in which the supermarket does not have sufficient self-generation of heat and needs to consume from the \gls{DHN}, behaving as a consumer in the market.
In this case, Figure \ref{fig:supermarket_dem} depicts the annual percentage of heat supplied by the heat producers to the supermarket. In general, the supermarket is mainly supplied by \gls{CHP} and the data center, since these agents have a large thermal capacity. As the supermarket is closer to the \gls{CHP}, when considering the distance criteria (\gls{P2P} Distance), the heat supplied by this resource, reaches its peak. Hence, as the data center is the farthest resource from the supermarket, the heat exchange reaches its low. The same line of thought is true for the \gls{P2P} Losses. Conversely, as the heat pump is the resource with the lowest $CO_2$ emissions, this resource reaches its maximum when considering the \gls{P2P} $CO_2$ market design.

\begin{figure}[!h]
        \begin{center}
        \includegraphics[width=0.50\textwidth]{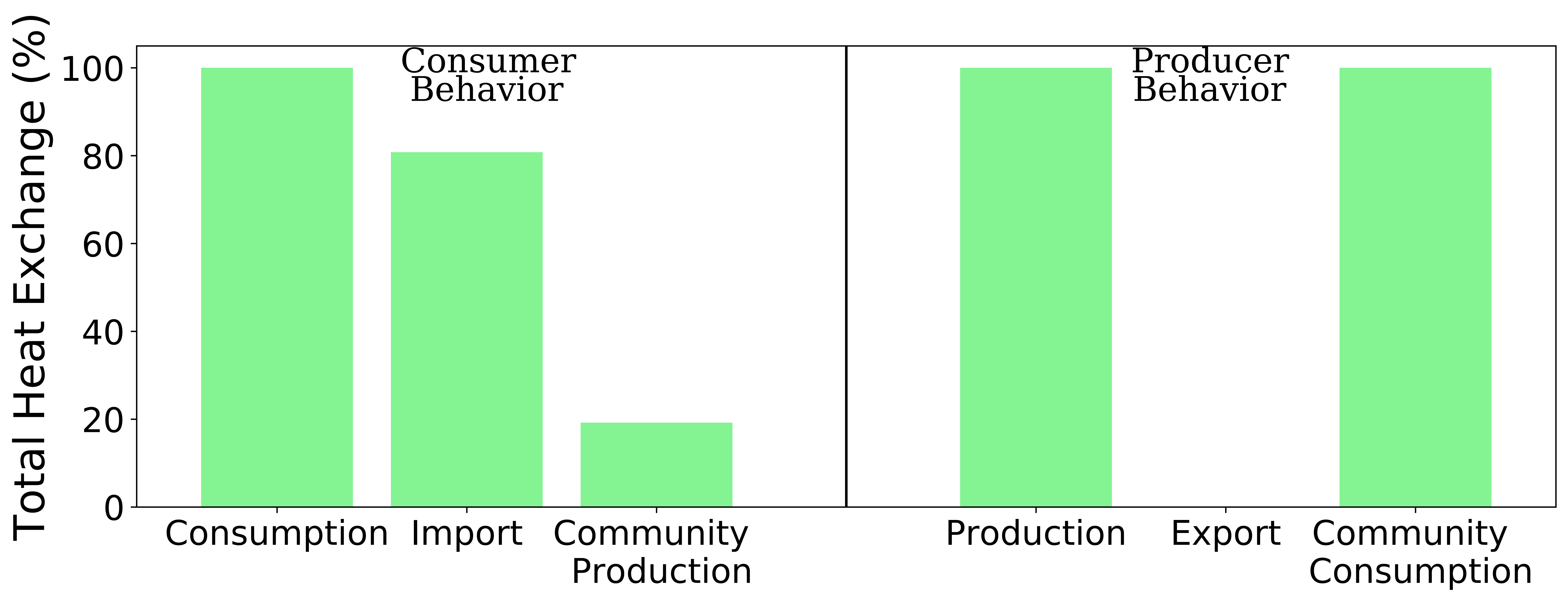}
        \caption{Supermarket heat exchange in the Community design.}
        \label{fig:supermarket_com}
    \end{center}
\end{figure}

Looking at the community-based market design (Figure \ref{fig:supermarket_com}), one can see that as a consumer, the supermarket is compelled to import about 80\% of the heat, the remaining 20\% being supplied by the community itself (heat pump). As a heat producer, all production is shared with the community itself, and no heat is exported.

\section{Conclusion}

District heating still has a long way to go, especially regarding the way heat is exchanged and the infrastructure needed for this transformation. Within this scope, new market models for district heating have been proposed in this work, encouraging direct heat exchange between peers. The network characteristics and impact on heat exchange were also assessed through product differentiation, giving to the peers and network operators the possibility to define and test criteria that best fits their interests. All markets designs were simulated, compared and incorporated in the market module of the EMB3Rs platform.

The results point to the feasible implementation of this type of market structure in \glspl{DHN}. The Full \gls{P2P} model presents the best results, since it disregards any limitations of the \gls{DHN} for the heat exchanges between the different players. This work, also proves that it is possible to impact the way heat is distributed according to preferences that may be associated with distance, minimizing losses or mitigating $CO_2$ emissions. As an example, analyzing the market design of \gls{P2P} Distance, one can see that the supermarket can increase by 500\% the heat supply to closest consumers when compared to the Full \gls{P2P} market design. In addition, the Community-based market design also reveals the possibility to divide agents into communities, allowing them to manage their own community and exchange heat with other communities, through heat import or export. Overall, if looking at the equilibrium between the agent participation in the market, the quality indicators do not show a balanced system. This is linked to the different heat technologies and prices, that change over the year according to several factors as the weather. The \gls{MiM} also highlights this point, as a low value for this indicator means a big difference between the maximum and minimum heat traded amongst the agents. 

Future work will focus on full network thermal characterization and comparison with the main findings here presented. Also larger networks will be explored in order to test the solutions in a real-world like environment.

\section*{Acknowledgements}
This work is supported by the European Union’s Horizon 2020 through the EU framework Program for Research and Innovation, within the EMB3Rs project under the agreement No. 847121.
In addition, we would like to thank Tiago Sousa for the insight comments that allowed us to improve the paper. 

\section*{Data Availability}
Datasets related to this article can be found at http://dx.doi.org/10.17632/ydbcpb73t2.1, an open-source online data repository hosted at Mendeley Data (António, Tiago, Zenaida, José, 2020).

\bibliographystyle{IEEEtran}
\bibliography{conference_101719}



\end{document}

%% file: Preliminaries.tex
\usepackage{graphicx} 
\usepackage{lineno,hyperref} 
\usepackage{xcolor}
\usepackage{amsmath,amssymb,amsfonts} 

\usepackage[acronym, automake]{glossaries} 

\makeglossaries
\newacronym{DHC}{DHC}{District Heating and Cooling}
\newacronym{DHN}{DHN}{Distric Heating Network}
\newacronym{IoT}{IoT}{Internet-of-Things}
\newacronym{P2P}{P2P}{Peer-to-Peer}
\newacronym{MV}{MV}{Medium-Voltage}
\newacronym{ADH}{ADH}{Average Dispatched Heat}
\newacronym{SPM}{SPM}{Successful Participation in the Market}
\newacronym{SM}{SM}{Supermarket}
\newacronym{DC}{DC}{Data Center}
\newacronym{HP}{HP}{Heat Pump}
\newacronym{COP}{COP}{Coefficient of Performance}
\newacronym{CHP}{CHP}{Combined Heat and Power}
\newacronym{QoS}{QoS}{Quality of Service}
\newacronym{QoE}{QoE}{Quality of Experience}
\newacronym{MiM}{MiM}{Min-Max Indicator}

%% file: Heat_P2P_markets.tex
\section{District Heating Market Designs}
The \gls{DHC} market designs discussed in this work, represent insights into the future of heat exchange in \gls{DHN}s. There is still a long way to go regarding infrastructure and legislation for the implementation of liberalized markets. In this context, the first steps in what we believe could be the \gls{DHC} systems of tomorrow are given in this work. In this way, pool, \gls{P2P} and community-based market approaches are addressed. Note that for the rest of the work, it is assumed that the heat sources are considered producers and the heat sinks are consumers.

\subsection{Pool Market Design}
The pool market designs consists of representing the merit order mechanism and obtain the market clearing price through the intersection of supply and demand curves. Thus, the market has the goal of maximizing the social welfare, meaning that lower offers from producers and higher offers from consumers are accepted. Mathematically, this market can be presented as:
\begin{subequations}\label{eq:pool_heat_math}
    \begin{align}
        \label{eq:pool_objective}
        \min_D \quad& \sum_{n \in \Omega_n} C_n P_n \\
        \label{eq:pool_agent_bounds}
        \text{s.t.} \quad&\underline{P}_{n} \leq P_{n} \leq  \overline{P}_{n} && p \in \Omega_n \\
        \label{eq:pool_balance}
        &\sum_{n \in \Omega_n} P_n  = 0 \\
        \label{eq:pool_consumer_neg}
        & P_n \leq 0 && n \in \Omega_c \\
        \label{eq:pool_producer_pos}
        & P_n \geq 0 && n \in \Omega_p
    \end{align}
\end{subequations}
where $D={\{P_n\in\mathbb{R}\}}_{n \in \Omega_n}$ correspond to the energy traded by each agent $n$.
$C_n$ represents the agents' bid price;  $\underline{P_n}$, $\overline{P_n}$,   represent the lower and upper bound of the agents' energy offer, respectively; $\Omega_c$ represent the consumers sets, $\Omega_p$ represent the producer sets. Eq \eqref{eq:pool_agent_bounds} set the agents offers boundaries. Eq \eqref{eq:pool_balance} sets the market balance, where the supply must equal the demand. \eqref{eq:pool_consumer_neg} sets that the consumption is non-positive in the system, while \eqref{eq:pool_producer_pos} sets that production variable from producers is non-negative. 

\subsection{P2P Market Design}

Regarding the P2P approach, it is proposed that two different peers can trade heat on a bilateral basis, without a third party supervision \cite{Sorin2019}. That is, each peer $n$ can exchange with another peer $m$ on an individual basis, defining the amount of energy to be bought or sold at a given price. This problem can be mathematically formulated as follows:

\begin{subequations}\label{eq:p2p_heat_math}
    \begin{align}
        \label{eq:p2p_objective}
        \min_D \quad& \sum_{n \in \Omega_n} C_n P_n &&\\
        \label{eq:p2p_balance}
        \text{s.t.} \quad& P_n=\sum_{m \in \Omega_n} P_{n,m} && n \in \Omega_n \\
        \label{eq:p2p_agents_bounds}
        & \underline{P}_{n} \leq P_{n} \leq  \overline{P}_{n} && n \in \Omega_n \\
        \label{eq:p2p_bidirectional}
        & P_{n,m}+P_{m,n}=0 && \{n, m\} \in \{\Omega_n\}\\
        \label{eq:p2p_consumer_neg}
        & P_n \leq 0 && n \in \Omega_c \\
        \label{eq:p2p_producer_pos}
        & P_n \geq 0 && n \in \Omega_p
    \end{align}
\end{subequations}
where $D={\{P_{n}\in\mathbb{R}\}}_{n \in \Omega_n}$ represents the heat traded by each agent $n$. Like in the pool market, the goal is to minimize the cost associated with the agents' transactions \eqref{eq:p2p_objective}. The total heat traded by an agent $n$ must equal the sum of the heat exchanges from that agent $n$ to the other agents $m$ \eqref{eq:p2p_balance}. Also, a reciprocity is expected in the bilateral trades \eqref{eq:p2p_bidirectional}, where $P_{n,m}$ and $P_{m,n}$ must be symmetric.

Looking at the peer-to-peer formulation, one can see that it yields the trade between agents. Thus, a preference can be added to each of these trades, which can be translated into a penalty or benefit. This is called product differentiation, meaning that a certain trade can be advantageous or harmful to the system management. In this way, the objective function is willing to benefit or penalize the trades that deserve such consideration. The distance between agents, the thermal losses and the CO$_{2}$ emissions are preferences that can be placed within this scope. There is also the option where the agents can choose the penalty that best suits their ideology. For instance, on the EMB3Rs platform, three different penalty options are provided to the consumers. One option is the physical network distance between agents. For example, an agent can select the distance penalty if he wishes to trade with the nearest neighbor. Another option is thermal losses, where an agent can select the thermal losses penalty if it is concerned about the system energy efficiency. Alternatively, the $CO_2$ penalty is proposed if an agent has environmental concerns. Conventionally, the product differentiation is represented as:
\begin{equation}
    C_{n,m}=P_{n,m}c_{n,m} 
    \label{eq:penalty}
\end{equation}
where $C_{n,m}$ represents the final penalty applied to the trade between agents $n$ and $m$. $P_{n,m}$ represents the energy trade between agents $n$ and $m$, and $c_{n,m}$ represents the initial penalty between these agents.

In order to apply product differentiation, the objective function must account with the penalty from Eq \eqref{eq:penalty}. Thus, the objective function takes the following form: 

\begin{equation}
    \min_{D} \sum_{n \in \Omega_n}C_nP_n + \sum_{n \in \Omega_n}\sum_{m \in \Omega_n} C_{n,m}
    \label{eq:prod_diff_obj}
\end{equation}
where $D=\{P_n,C_{n,m}\}\in\mathbb{R}_{n,m \in \Omega_n}$.
Hence, the formulation is completed, since equations \eqref{eq:p2p_balance}-\eqref{eq:p2p_producer_pos} keep unchanged. Nevertheless, the determination of the product differentiation penalties may follow different ways.

\subsubsection{Physical Network Distance Preference}
In the distance preference, the network distance between the selected agents is determined. The penalty implies the sum up of all the pipes that make the path between agents. Note that Dijkstra's algorithm \cite{dijkstra1959note} is used to find the shortest path between agents. Thus, the penalty associated to the network distance is given by:    

\begin{equation}
        c_{n,m}=\sum_{i \in \Omega_{I_{ n,m}}}d_{i,n,m}/Tot Dist
        \label{eq:distance}
\end{equation}
where $d_{i,n,m}$ represents the pipe distance along the  path between agents $n$ and $m$, while 385.08m is the total network distance.

\subsubsection{Network Thermal Losses Preference}

The thermal losses penalty between two agents is given by the share that each agent has in the system losses considering the thermal flow in each pipe. In this case, it is required to determine the thermal flow in the \gls{DHN} and, therefore, the losses in each pipe. To determine the thermal flow and losses in the \gls{DHN} based on the initial market results, the thermal control algorithm in \cite{Cao2018} is used. Therefore, the impact that each agent has on the thermal flow and losses of each pipeline is determined using Bialek's downstream looking algorithm \cite{Bialek1996}. Finally, the thermal losses penalty for the transaction between two peers is given by:     
    
\begin{equation}
        c_{n,m}=\sum_{i \in \Omega_{I_{ n,m}}}l_{i,n,m} D_{i,n,m} d_{i,n,m}/Tot Loss
        \label{eq:losses}
\end{equation}
where $l_{i,n,m}$ represents the thermal losses in each pipe along the path between agents $n$ and $m$; $D_{i,n,m}$ represents the $n,m$ peer impact in each pipe of the system determined by the downstream looking algorithm presented in \cite{Bialek1996}. In this way, a fairly penalty allocation for the transaction between two agents is achieved, accounting for the cumulative impact that such transaction has in the thermal losses in the system.
    
\subsubsection{$CO_2$ Emissions Preference}
The last option proposed for product differentiation is to penalize transactions through $CO_2$ emissions. This penalty consists of penalizing peer transactions that may, consequently, emit higher emissions into the atmosphere. The EMB3Rs platform can provide standard levels of $CO_2$ per technology, and therefore, penalties between agents $n$ and $m$ consider such levels. Here, the penalty is only associated with the heat source. Hence, the $CO_2$ penalty between agents $n$ and $m$ is given by the quotient between agent $n$ emissions and the total system emissions:   
    
        \begin{equation}
        c_{n,m}=E_n/\sum_{n \in \Omega_n} E_n
        \label{eq:co2}
    \end{equation}
where $E_n$ represents the $CO_2$ emissions by agent $n$.
\subsection{Community-based Market}

The community-based market design intends to represent a more hierarchical structure of bilateral peer trades. In general, a community is composed by members who share common interests or are geographically close. In this model, there is a community manager responsible for the community's energy management. This manager supervises all the trading activities within the community, as well as works as an intermediary in the heat trade with other communities or with the main grid \cite{Sousa2019}. The mathematical formulation is presented as:

\begin{subequations}\label{eq:p2p_heat_math}
    \begin{align}
        \begin{aligned}
        \label{eq:community_objective}
        \min_{D}  \sum_{n\in\Omega_n} \sum_{k \in \Omega_{k}} C_{n,k}P_{n,k}-c_{exp,k}q_{exp,k} \\
        +c_{imp,k}q_{imp,k} \\
        \end{aligned} 
        \\
        \begin{aligned}
        \label{com_bidirectional}
        P_{k,k'}+P_{k',k}=0, \forall (k,k') \in (\Omega_{k})\\
        \end{aligned}
        \\
        \begin{aligned}
        \label{export_balance}
        q_{exp,k'}=\sum_{k\in\Omega_{k}} P_{k',k}, \forall k' \in \Omega_{k} 
        \end{aligned}
        \\
        \begin{aligned}
        \label{import_balance}
        q_{imp,k'}=\sum_{k\in\Omega_{k}} P_{k',k}, \forall k' \in \Omega_{k} \\
        \end{aligned}
        \\
        \begin{aligned}
        \label{com_bilateral_trades}
        \sum_{k \in \Omega_{k}}P_{k',k}=q_{exp,k'}-q_{imp,k'}, \\
        \forall k' \in \Omega{k} \\
        \end{aligned}
        \\
         \begin{aligned}
        \label{com_agent_balance}
        P_{n,k}+q_{n,k}+\alpha_{n,k}-\beta_{n,k} = 0, \\
        \forall (n,k) \in (\Omega_n, \Omega_{k})   \\
        \end{aligned}
        \\
        \begin{aligned}
        \label{com_balance}
        \sum_{n\in\Omega_n} q_{n,k} = 0, \forall k \in \Omega_{k} \\
        \end{aligned}
        \\
        \begin{aligned}
        \label{com_export_balance}
        \sum_{n\in\Omega_n} \beta_{n,k} = q_{exp,k}, \forall k \in \Omega_{k} \\
        \end{aligned}
        \\
        \begin{aligned}
        \label{com_import_balance}
        \sum_{n\in\Omega_n} \alpha_{n,k} = q_{imp,k}, \forall k \in \Omega_{k} \\ 
        \end{aligned}
        \\
        \begin{aligned}
        \label{com_agents_limits}
        \underline{P}_{n} \leq P_{n} \leq  \overline{P}_{n} && (n,k) \in (\Omega_n,\omega_{k})
        \end{aligned}
    \end{align}
\end{subequations}
where $D={\{P_{n,k}, q_{exp,k}, q_{imp,k}\in\mathbb{R}\}}_{(n,k) \in (\Omega_n,\Omega{k}}$. $P_{n,k}$ represents the internal trade of agent $n$ within its own community $k$. \eqref{com_bidirectional} represents the symmetry when communities exchange heat. Equation \eqref{export_balance} balances the exported heat by a community with other communities. The same is valid for \eqref{import_balance}, regarding the imported heat. Also, the sum of one community bilateral trades must equal the exported heat minus the imported heat \eqref{com_bilateral_trades}. Equation \eqref{com_agent_balance} sets agents' balance, i.e., the purchase/consumption, the heat traded within the community and the heat exchanged with other communities must reach an equilibrium in each time period. Within a community, the purchase/consumption of all involved agents must be equal to zero \eqref{com_balance}. Furthermore, the heat exported by each community agent must equal the total heat exported by the community \eqref{com_export_balance}. The same holds true for the imported heat \eqref{com_import_balance}. Like in the previous market designs, heat boundaries ought to be kept \eqref{com_agents_limits}.

%% file: conference_101719.bbl
\begin{thebibliography}{10}
\providecommand{\url}[1]{#1}
\csname url@samestyle\endcsname
\providecommand{\newblock}{\relax}
\providecommand{\bibinfo}[2]{#2}
\providecommand{\BIBentrySTDinterwordspacing}{\spaceskip=0pt\relax}
\providecommand{\BIBentryALTinterwordstretchfactor}{4}
\providecommand{\BIBentryALTinterwordspacing}{\spaceskip=\fontdimen2\font plus
\BIBentryALTinterwordstretchfactor\fontdimen3\font minus
  \fontdimen4\font\relax}
\providecommand{\BIBforeignlanguage}[2]{{%
\expandafter\ifx\csname l@#1\endcsname\relax
\typeout{** WARNING: IEEEtran.bst: No hyphenation pattern has been}%
\typeout{** loaded for the language `#1'. Using the pattern for}%
\typeout{** the default language instead.}%
\else
\language=\csname l@#1\endcsname
\fi
#2}}
\providecommand{\BIBdecl}{\relax}
\BIBdecl

\bibitem{Mathiesen2019}
\BIBentryALTinterwordspacing
B.~V. Mathiesen, N.~Bertelsen, N.~C.~A. Schneider, L.~S. García,
  S.~Paardekooper, J.~Z. Thellufsen, and S.~R. Dj{\o}rup, ``{Towards a
  decarbonised heating and cooling sector in Europe: Unlocking the potential of
  energy efficiency and district energy},'' \emph{Aalborg Universitet}, p.~98,
  2019. [Online]. Available:
  \url{https://vbn.aau.dk/ws/portalfiles/portal/316535596/Towards_a_decarbonised_H_C_sector_in_EU_Final_Report.pdf}
\BIBentrySTDinterwordspacing

\bibitem{Persson2011}
U.~Persson and S.~Werner, ``{Heat distribution and the future competitiveness
  of district heating},'' \emph{Applied Energy}, vol.~88, no.~3, pp. 568--576,
  2011.

\bibitem{Buffa2019}
S.~Buffa, M.~Cozzini, M.~D'Antoni, M.~Baratieri, and R.~Fedrizzi, ``{5th
  generation district heating and cooling systems: A review of existing cases
  in Europe},'' \emph{Renewable and Sustainable Energy Reviews}, vol. 104, no.
  October 2018, pp. 504--522, 2019.

\bibitem{Magnusson2016}
D.~Magnusson, ``{Who brings the heat? – From municipal to diversified
  ownership in the Swedish district heating market post-liberalization},''
  \emph{Energy Research {\&} Social Science}, vol.~22, pp. 198--209, dec 2016.

\bibitem{Westin2002}
P.~Westin and F.~Lagergren, ``{Re-regulating district heating in Sweden},''
  \emph{Energy Policy}, vol.~30, no.~7, pp. 583--596, 2002.

\bibitem{Soderholm2011}
P.~S{\"{o}}derholm and L.~W{\aa}rell, ``{Market opening and third party access
  in district heating networks},'' \emph{Energy Policy}, vol.~39, no.~2, pp.
  742--752, 2011.

\bibitem{Liu2019}
W.~Liu, D.~Klip, W.~Zappa, S.~Jelles, G.~J. Kramer, and M.~van~den Broek,
  ``{The marginal-cost pricing for a competitive wholesale district heating
  market: A case study in the Netherlands},'' \emph{Energy}, vol. 189, p.
  116367, 2019.

\bibitem{Zhang2013}
J.~Zhang, B.~Ge, and H.~Xu, ``{An equivalent marginal cost-pricing model for
  the district heating market},'' \emph{Energy Policy}, vol.~63, pp.
  1224--1232, 2013.

\bibitem{Wissner2014}
M.~Wissner, ``{Regulation of district-heating systems},'' \emph{Utilities
  Policy}, vol.~31, pp. 63--73, 2014.

\bibitem{Gebremedhin2004}
A.~Gebremedhin and B.~Moshfegh, ``{Modelling and optimization of district
  heating and industrial energy system - An approach to a locally deregulated
  heat market},'' \emph{International Journal of Energy Research}, vol.~28,
  no.~5, pp. 411--422, 2004.

\bibitem{Gulzar2015}
K.~Gulzar, S.~Sierla, V.~Vyatkin, N.~Papakonstantinou, P.~G. Flikkema, and
  C.~W. Yang, ``{An auction-based smart district heating grid},'' \emph{IEEE
  International Conference on Emerging Technologies and Factory Automation,
  ETFA}, pp. 1--8, 2015, {Luxembourg}.

\bibitem{Burger2019}
V.~B{\"{u}}rger, J.~Steinbach, L.~Kranzl, and A.~M{\"{u}}ller, ``{Third party
  access to district heating systems - Challenges for the practical
  implementation},'' \emph{Energy Policy}, vol. 132, no. October 2018, pp.
  881--892, 2019.

\bibitem{Marinova2008}
M.~Marinova, C.~Beaudry, A.~Taoussi, M.~Tr{\'{e}}panier, and J.~Paris,
  ``{Economic assessment of rural district heating by bio-steam supplied by a
  paper mill in Canada},'' \emph{Bulletin of Science, Technology {\&} Society},
  vol.~28, no.~2, pp. 159--173, 2008.

\bibitem{Syri2015}
S.~Syri, H.~M{\"{a}}kel{\"{a}}, S.~Rinne, and N.~Wirgentius, ``{Open district
  heating for Espoo city with marginal cost based pricing},'' \emph{12th
  International Conference on the European Energy Market, EEM}, pp. 1--5, 2015,
  {Lisbon}, Portugal.

\bibitem{Brand2014}
L.~Brand, A.~Calv{\'{e}}n, J.~Englund, H.~Landersj{\"{o}}, and P.~Lauenburg,
  ``{Smart district heating networks - A simulation study of prosumers' impact
  on technical parameters in distribution networks},'' \emph{Applied Energy},
  vol. 129, pp. 39--48, 2014.

\bibitem{Liu2016}
X.~Liu and P.~Mancarella, ``{Modelling, assessment and Sankey diagrams of
  integrated electricity-heat-gas networks in multi-vector district energy
  systems},'' \emph{Applied Energy}, vol. 167, pp. 336--352, 2016.

\bibitem{Huang2017}
J.~Huang, Z.~Li, and Q.~H. Wu, ``{Coordinated dispatch of electric power and
  district heating networks: A decentralized solution using optimality
  condition decomposition},'' \emph{Applied Energy}, vol. 206, no. September,
  pp. 1508--1522, 2017.

\bibitem{Lu2018}
S.~Lu, W.~Gu, J.~Zhou, X.~Zhang, and C.~Wu, ``{Coordinated dispatch of
  multi-energy system with district heating network: Modeling and solution
  strategy},'' \emph{Energy}, vol. 152, pp. 358--370, 2018.

\bibitem{Dominkovic2018}
D.~F. Dominkovi{\'{c}}, M.~Wahlroos, S.~Syri, and A.~S. Pedersen, ``{Influence
  of different technologies on dynamic pricing in district heating systems:
  Comparative case studies},'' \emph{Energy}, vol. 153, no. March, pp.
  136--148, 2018.

\bibitem{Li2019}
H.~Li, J.~Song, Q.~Sun, F.~Wallin, and Q.~Zhang, ``{A dynamic price model based
  on levelized cost for district heating},'' \emph{Energy, Ecology and
  Environment}, vol.~4, pp. 15--25, 2019.

\bibitem{Djorup2020}
S.~Djørup, K.~Sperling, S.~Nielsen, P.~A. Østergaard, J.~Z. Thellufsen,
  P.~Sorknæs, H.~Lund, and D.~Drysdale, ``{District heating tariffs, economic
  optimisation and local strategies during radical technological change},''
  \emph{Energies}, vol.~13, no.~5, p. 1172, 2020.

\bibitem{Bhattacharya2016}
S.~Bhattacharya, V.~Chandan, V.~Arya, and K.~Kar, ``{Fairness based demand
  response in DHC networks using real time parameter identification},''
  \emph{2016 IEEE International Conference on Smart Grid Communications,
  SmartGridComm 2016}, pp. 32--37, 2016.

\bibitem{Pazeraite2013}
A.~Pažėraitė and M.~Krakauskas, ``{Towards liberalized district heating
  market : Kaunas city case},'' \emph{Management of Organizations: Systematic
  Research}, vol.~67, pp. 53--67, 2013.

\bibitem{opendistrict}
\BIBentryALTinterwordspacing
Exergi. Open district heating. [Online]. Available:
  \url{https://www.opendistrictheating.com/}
\BIBentrySTDinterwordspacing

\bibitem{Li2015}
H.~Li, Q.~Sun, Q.~Zhang, and F.~Wallin, ``{A review of the pricing mechanisms
  for district heating systems},'' \emph{Renewable and Sustainable Energy
  Reviews}, vol.~42, pp. 56--65, 2015.

\bibitem{Moshkin2016}
I.~Moshkin and A.~Sauhats, ``{Solving district heating optimization problems in
  the market conditions},'' \emph{57th International Scientific Conference on
  Power and Electrical Engineering of Riga Technical University (RTUCON)}, pp.
  1--6, 2016, {Riga}, Latvia.

\bibitem{Valeriy2019}
D.~Valeriy and K.~Dmytro, ``{Functional structure of the local thermal energy
  market in district heating},'' \emph{2019 IEEE 6th International Conference
  on Energy Smart Systems, ESS 2019 - Proceedings}, vol.~1, pp. 343--346, 2019.

\bibitem{Karlsson2009}
M.~Karlsson, A.~Gebremedhin, S.~Klugman, D.~Henning, and B.~Moshfegh,
  ``{Regional energy system optimization - Potential for a regional heat
  market},'' \emph{Applied Energy}, vol.~86, no.~4, pp. 441--451, 2009.

\bibitem{embers}
\BIBentryALTinterwordspacing
EMB3Rs. {EMB3Rs} - heat and cold matching platform. [Online]. Available:
  \url{https://www.emb3rs.eu/}
\BIBentrySTDinterwordspacing

\bibitem{climatex}
\BIBentryALTinterwordspacing
S.~Donnellan, F.~Burns, O.~Alabi, and R.~Low, ``Lessons from european
  regulation and practice for scottish district heating regulation,''
  \emph{ClimateXChange}, 2018. [Online]. Available:
  \url{https://www.climatexchange.org.uk/media/3569/lessons-from-european-district-heating-regulation.pdf}
\BIBentrySTDinterwordspacing

\bibitem{DanishEnergyAgency2016}
\BIBentryALTinterwordspacing
{Danish Energy Agency}, ``{Regulation and planning of district heating in
  Denmark},'' pp. 1--27, 2016. [Online]. Available:
  \url{https://ens.dk/sites/ens.dk/files/contents/material/file/regulation_and_planning_of_district_heating_in_denmark.pdf}
\BIBentrySTDinterwordspacing

\bibitem{Aanensen2014}
\BIBentryALTinterwordspacing
T.~Aanensen and N.~Fedoryshyn, ``District heating and district cooling in
  norway,'' \emph{Statistics Norway}, 2014. [Online]. Available:
  \url{https://www.ssb.no/en/energi-og-industri/artikler-og-publikasjoner/_attachment/184839?_ts=1475e7199a8}
\BIBentrySTDinterwordspacing

\bibitem{Hawkey2014}
D.~Hawkey and J.~Webb, ``District energy development in liberalised markets:
  situating uk heat network development in comparison with dutch and norwegian
  case studies,'' \emph{Technology Analysis \& Strategic Management}, vol.~26,
  no.~10, pp. 1228--1241, 2014.

\bibitem{Aberg2016}
M.~{\AA}berg, L.~F{\"{a}}lting, and A.~Forssell, ``{Is Swedish district heating
  operating on an integrated market? - Differences in pricing, price
  convergence, and marketing strategy between public and private district
  heating companies},'' \emph{Energy Policy}, vol.~90, pp. 222--232, 2016.

\bibitem{Paiho2016}
S.~Paiho and F.~Reda, ``{Towards next generation district heating in
  Finland},'' \emph{Renewable and Sustainable Energy Reviews}, vol.~65, pp.
  915--924, 2016.

\bibitem{Werner2017}
S.~Werner, ``{International review of district heating and cooling},''
  \emph{Energy}, vol. 137, pp. 617--631, 2017.

\bibitem{Sorknaes2020}
P.~Sorkn{\ae}s, H.~Lund, I.~R. Skov, S.~Dj{\o}rup, K.~Skytte, P.~E. Morthorst,
  and F.~Fausto, ``{Smart energy markets - future electricity, gas and heating
  markets},'' \emph{Renewable and Sustainable Energy Reviews}, vol. 119, no.
  March, p. 109655, 2020.

\bibitem{emb3rs_2}
\BIBentryALTinterwordspacing
EMB3Rs, ``User-driven energy-matching \& business prospection tool for
  industrial excess heat/cold reduction, recovery and redistribution,''
  {Horizon} 2020, Nº 847121. [Online]. Available:
  \url{https://cordis.europa.eu/project/id/847121}
\BIBentrySTDinterwordspacing

\bibitem{Sorin2019}
E.~Sorin, L.~Bobo, and P.~Pinson, ``{Consensus-based approach to peer-to-peer
  electricity markets with product differentiation},'' \emph{IEEE Transactions
  on Power Systems}, vol.~34, no.~2, pp. 994--1004, 2019.

\bibitem{dijkstra1959note}
E.~W. Dijkstra, ``A note on two problems in connexion with graphs,''
  \emph{Numerische mathematik}, vol.~1, no.~1, pp. 269--271, 1959.

\bibitem{Cao2018}
Y.~Cao, W.~Wei, L.~Wu, S.~Mei, M.~Shahidehpour, and Z.~Li, ``{Decentralized
  operation of interdependent power distribution network and district heating
  network: a market-driven approach},'' \emph{IEEE Transactions on Smart Grid},
  vol.~PP, no.~c, p.~1, 2018.

\bibitem{Bialek1996}
J.~Bialek, ``{Tracing the flow of electricity},'' \emph{IEE Proceedings:
  Generation, Transmission and Distribution}, vol. 143, no.~4, pp. 313--320,
  1996.

\bibitem{Sousa2019}
T.~Sousa, T.~Soares, P.~Pinson, F.~Moret, T.~Baroche, and E.~Sorin,
  ``{Peer-to-peer and community-based markets: A comprehensive review},''
  \emph{Renewable and Sustainable Energy Reviews}, vol. 104, no. January, pp.
  367--378, 2019.

\bibitem{dataset_men}
A.~S. Faria, T.~Soares, J.~M. Cunha, and Z.~Mourão. (2020)
  \BIBforeignlanguage{English}{District heating database}. \url
  http://dx.doi.org/10.17632/ydbcpb73t2.1.

\bibitem{Gadd2014}
H.~Gadd and S.~Werner, ``{Achieving low return temperatures from district
  heating substations},'' \emph{Applied Energy}, vol. 136, pp. 59--67, 2014.

\bibitem{BrandThesis2014}
M.~Brand, ``{Heating and domestic hot water systems in buildings supplied by
  low-temperature district heating},'' \emph{Technical University of Denmark},
  pp. 1--141, 2014.

\bibitem{hofor}
\BIBentryALTinterwordspacing
HOFOR. The price for district heating 2020 for business customers. [Online].
  Available:
  \url{https://www.hofor.dk/erhverv/priser-paa-forsyninger-erhvervskunder/prisen-fjernvarme-2020-erhvervskunder/}
\BIBentrySTDinterwordspacing

\bibitem{Ommen2013}
T.~Ommen, W.~B. Markussen, and B.~Elmegaard, ``{Heat pumps in district heating
  networks},'' \emph{2nd Symposium on Advances in Refrigeration and Heat Pump
  Technology}, pp. 1--9, 2013, {Odense}, Denmark.

\bibitem{Fukano2011}
S.~Fukano, T.~Kudo, and N.~Arata, ``{Ammonia heat pump package using waste heat
  as source},'' \emph{4th IIR International Conference on Ammonia Refrigeration
  Technology}, pp. 2946--2951, 2011, {Ohrid}, Macedonia.

\bibitem{energi}
\BIBentryALTinterwordspacing
ENERGI. Welcome to energi data service. [Online]. Available:
  \url{https://www.energidataservice.dk/}
\BIBentrySTDinterwordspacing

\bibitem{datacenter}
\BIBentryALTinterwordspacing
M.~Monroe. How to reuse waste heat from data centers intelligently. [Online].
  Available:
  \url{https://www.datacenterknowledge.com/archives/2016/05/10/how-to-reuse-waste-heat-from-data-centers-intelligently}
\BIBentrySTDinterwordspacing

\bibitem{WAHLROOS20171228}
M.~Wahlroos, M.~Pärssinen, J.~Manner, and S.~Syri, ``Utilizing data center
  waste heat in district heating – impacts on energy efficiency and prospects
  for low-temperature district heating networks,'' \emph{Energy}, vol. 140, pp.
  1228 -- 1238, 2017.

\bibitem{henry_hub}
\BIBentryALTinterwordspacing
{U.S. Energy Information}. Natural gas. [Online]. Available:
  \url{https://www.eia.gov/dnav/ng/hist/rngwhhdd.htm}
\BIBentrySTDinterwordspacing

\bibitem{weather}
\BIBentryALTinterwordspacing
rp5u. Weather archive in copenhagen. [Online]. Available:
  \url{https://rp5.ru/Weather_archive_in_Copenhagen,_Kastrup_(airport),_METAR}
\BIBentrySTDinterwordspacing

\bibitem{Karampour2014}
\BIBentryALTinterwordspacing
M.~Karampour and S.~Sawalha, ``{Supermarket refrigeration and heat recovery
  using $CO_2$ as refrigerant: A comprehensive evaluation based on field
  measurements and modelling},'' no. June, p.~45, 2014. [Online]. Available:
  \url{https://www.diva-portal.org/smash/get/diva2:849667/FULLTEXT01.pdf}
\BIBentrySTDinterwordspacing

\bibitem{Id2018}
L.~R. Adrianto, P.-A. Grandjean, and S.~Sawalha, ``{Heat recovery from $CO_2$
  refrigeration system in supermarkets to district heating network},''
  \emph{13th IIR Gustav Lorentzen Conference on Natural Refrigerants (GL2018)},
  pp. 1--8, 2018, {Valencia}, Spain.

\bibitem{thermos}
\BIBentryALTinterwordspacing
THERMOS. Thermal energy resource modelling and optimization system. [Online].
  Available: \url{https://www.thermos-project.eu/home/}
\BIBentrySTDinterwordspacing

\bibitem{EIB2018}
\BIBentryALTinterwordspacing
{European Investment Bank}, \emph{{EIB project carbon footprint methodologies:
  Methodologies for the assessment of project GHG emissions and emission
  variations}}, 2020, no. July. [Online]. Available:
  \url{https://www.eib.org/attachments/strategies/eib_project_carbon_footprint_methodologies_en.pdf}
\BIBentrySTDinterwordspacing

\bibitem{co2_intensity}
\BIBentryALTinterwordspacing
{European Environment Agency}. Data visualization - $co_2$ emission intensity.
  [Online]. Available:
  \url{https://www.eea.europa.eu/data-and-maps/daviz/co2-emission-intensity-5#tab-googlechartid_chart_11_filters=%7B%22rowFilters%22%3A%7B%7D%3B%22columnFilters%22%3A%7B%22pre_config_ugeo%22%3A%5B%22European%20Union%20(current%20composition)%22%5D%7D%7D)}
\BIBentrySTDinterwordspacing

\bibitem{Moret2019}
F.~Moret and P.~Pinson, ``{Energy collectives: A community and fairness based
  approach to future electricity markets},'' \emph{IEEE Transactions on Power
  Systems}, vol.~34, no.~5, pp. 3994--4004, 2019.

\bibitem{Jain1998}
\BIBentryALTinterwordspacing
R.~Jain, D.~Chiu, and W.~Hawe, ``{A quantitative measure of fairness and
  discrimination for resource allocation in shared computer systems},''
  \emph{arXiv: Networking and Internet Architecture}, 1998. [Online].
  Available: \url{http://arxiv.org/abs/cs/9809099}
\BIBentrySTDinterwordspacing

\end{thebibliography}
